\documentclass[usenatbib,useAMS]{mn2e}

\usepackage{amssymb}
\usepackage{graphicx}

\title[Runaway Stars]{High Galactic latitude runaway stars as tracers of the spiral arms}
\author[M. D. V. Silva and R. Napiwotzki]{M.~D.~V.~Silva$^1$\thanks{E-mail: madusilva@gmail.com} and R.~Napiwotzki$^1$\\
$^1$Centre for Astrophysics Research, STRI, University of Hertfordshire, College Lane, Hatfield AL10 9AB}

\date{2013 January}

\pagerange{\pageref{firstpage}--\pageref{lastpage}} \pubyear{2012}

\begin{document}

\maketitle

\begin{abstract}
A direct observation of the spiral structure of the Galaxy is hindered by our position in the middle of the Galactic plane. We propose a method based on the analysis of the birthplaces of high Galactic latitude runaway stars to map the spiral arms and determine their dynamics. As a proof of concept, the method is applied to a local sample of early-type stars and a sample of runaways stars to obtain estimates of the pattern speed ($\Omega_{\mathrm{p},\mathrm{local}}=20.3\pm 0.5\ \mathrm{km}\, \mathrm{s}^{-1}\, \mathrm{kpc}^{-1}$ and $\Omega_{\mathrm{p},\mathrm{runaway}}=21.9\pm 8.6\ \mathrm{km}\, \mathrm{s}^{-1}\, \mathrm{kpc}^{-1}$) and the spiral arm's phase angle ($\phi_{\mathrm{local}}=-53.1^{\circ}\pm 3.1^{\circ}$ and $\phi_{\mathrm{runaway}}=-63.5^{\circ}\pm 22.5^{\circ}$). We also estimate the performance of this method once the data gathered by {\it Gaia}, in particular for runaway stars observed on the other side of the Galaxy, is available.
\end{abstract}


\section{Introduction}
\label{s:intro}
The spiral arms are readily visible in photographic images of face-on spiral galaxies as exemplified by the galaxy M51. They are usually traced, in visible light, by young hot stars and respective $\mathrm{H}\, \mathrm{II}$ regions or, in other words, are the preferred sites of star formation. Our position within the Galactic disc does not permit a direct observation of the spiral arms as they appear superimposed along our line of sight and are heavily obscured by dust. Nevertheless, many attempts have been made to determine their precise geometry and dynamics. The results obtained in different studies of the spiral structure of our Galaxy, using a variety of tracers (gas, dust and stars), were summarised and synthesised in a metastudy by \citet{vallee3}. The picture that emerges is that of a 4-arm spiral structure with pitch angle $p=12.8^{\circ}$. However, some controversy remains regarding the spiral structure properties, in particular the number of arms and their shape (pitch angle): \citet{lepine1}, for example, claim the existence of ``square''-shaped spiral arms, corresponding to the 4:1 resonance, instead of the usual logarithmic spiral arms. In any instance, only nearby segments of the spiral arms appear to be well defined in most studies (e.g. Fig.~3 of \citealp{russeil} and Fig.~4 of \citealp{lepine1}), whereas the spiral structure behind the Galactic centre is virtually unknown ({\it Zona Galactica Incognita} according to \citealp{vallee2}).

The determination of the spiral pattern speed $\Omega_\mathrm{p}$ is a closely related problem that has been the subject of many different studies. The different methods used to estimate the value of $\Omega_p$ are reviewed by \citet{gerhard1}. The most direct method consists in finding the birthplaces of open clusters or individual young stars by computing their orbits (e.g. \citealp{amaral1}; \citealp*{fernandez1}; \citealp{dias1}; \citealp{naoz1}).

In this paper we propose the use of a different tracer of the spiral arms structure: the Galactic population of high Galactic latitude runaway stars. It has already been shown that high quality astrometry, in particular that obtained with {\it Hipparcos} (see \citealp*{hoogerwerf}), permits the recovery of the birthplaces of runaway stars thus, the higher quality promised by the {\it Gaia} mission, will allow the analysis of a much larger volume of the Galaxy. Although it is more common to use larger structures as tracers, as the aforementioned $\mathrm{H}\, \mathrm{II}$ regions \citep{russeil} or open clusters \citep{dias1}, single stars were also previously used (e.g. Chepeid stars by \citealp*{majaess1} and O-B stars by \citealp{fernandez1}). However, ``normal'' early-type stars are confined to the disc and so are heavily affected by interstellar reddening, whereas high Galactic latitude runaway stars are not. Moreover, since runaway stars can travel large distances (several $\mathrm{kpc}$) away from their birthplaces, it is possible to map portions of the spiral arms that are further away than the runaway stars. 

Nevertheless, the number of high Galactic latitude runaway stars and the spatial distribution of their birthplaces does not permit yet the application of this method with high accuracy, however  the number of known stars of this type will increase by at least one order of magnitude with the {\it Gaia} satellite mission, which will deliver high accuracy astrometric data. Thus, in this paper our objective is to conduct a proof of concept study to show how the use of the birthplaces of runaway stars may be used to trace the spiral arms so it can be applied to the larger population of runaway stars observed by {\it Gaia}.

In order to accomplish this objective we apply a variation of the method used in other studies (usually with open clusters as tracers, e.g. \citealp{amaral1} and \citealp{dias1}). This method consists in using kinematical information about an object to compute its orbit in the Galaxy's potential and thus its birthplace (given an age estimate). Since objects with different ages indicate the position of the spiral arms in different time instants, this method also provides an estimate of the pattern speed $\Omega_\mathrm{p}$. Because of the shortcomings of the runaway stars sample we apply this method to a sample of local early-type stars as it provides a more accurate test (since it is larger and we may use the more accurate {\it Hipparcos} astrometry). After showing the results obtained with these two samples (local and runaways), we discuss the expected improvement in performance when this method is applied to {\it Gaia} data.
\section{Pattern speed determination}
\label{s:pattern}
\subsection{Samples}
\label{s:samples}
\begin{figure}
\includegraphics[width=82mm]{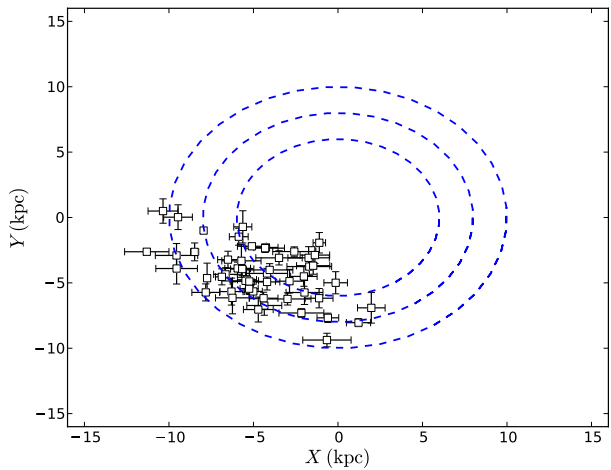}
\caption{Birthplaces of the runaway stars in the Galactic plane (frame centered in the centre of the Galaxy) including $1\, \sigma$ error bars. The concentric circles indicate distances of 5, 8 and $10\ \mathrm{kpc}$ to the centre.}
\label{f:1}
\end{figure}

The sample of runaway stars under consideration in this paper is the sample presented in \citet*{silva1}. This is a sample of 96 stars covering the brightness range $6.5<V<14.5$, and an altitude above the Galactic plane of $0.3 - 30.5\, \mathrm{kpc}$. The ejection velocities, flight times and birthplaces were computed as detailed in \citet{silva1}. In summary, the orbits were integrated backwards in an axisymmetric gravitational potential. Thus, the influence of the slight overdensities in the region of the spiral arms (less than 10~per~cent of the total gravitational potential, according to \citealp{fernandez1}),and the Galactic bar were ignored. This is justified for the following reasons: the total effect of the spiral arms in the velocity field is estimated to be only about $3\ \mathrm{km}\, \mathrm{s}^{-1}$, according to \citet{fernandez1}, which corresponds to a maximum systematic effect in the position of the birthplaces of $\sim 600\ \mathrm{pc}$ for the older stars (but less for the younger ones); this method is meant to be used with high Galactic latitude runaway stars and it was already shown in \citet{silva1} that these stars are not very sensitive to changes in the disc Galactic potential, given that they leave the Galactic plane with large ejection velocities. The 51 birthplaces which were determined with a precision $<1.5\, \mathrm{kpc}$ in both $x$ and $y$ directions are shown in Fig.~\ref{f:1}.

In principle it would be possible to fit directly these 51 points to determine the feasibility of the proposed method to determine the spiral arms shape, position and pattern speed, however this sample is small and the distances and proper motions have relatively large measurements errors (no parallaxes). For this reason a second sample, consisting of local early-type main sequence stars, was selected allowing for a better determination of birthplaces. This sample and the runaway stars sample will be referred to as sample~A and sample~B, respectively.

\begin{figure}%
\includegraphics[width=82mm]{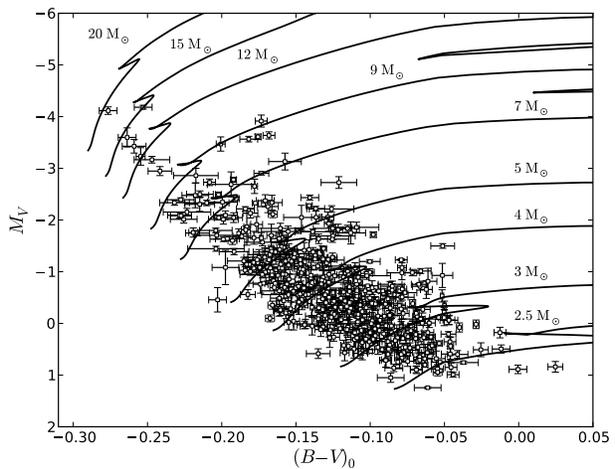}%
\caption{Deredenned colour-magnitude diagram of local early-type stars (sample~A) with respective error bars. The evolutionary tracks obtained through the bolometric corrections of \citet{flower1} from the theoretical tracks by \citet{ekstrom} are also plotted.}%
\label{f:2}%
\end{figure}

Sample~A was selected from a cross-match of the {\it Hipparcos} (new reduction by \citealp*{leeuwen1}) and the Bright Star \citep*{hoffleit1} catalogues. The selection criteria was: $B-V$ colour $<0.05$ (corresponding to an effective temperature $<9500\, \mathrm{K}$, cf. \citealp*{napiwotzki}), and parallax $>5\, \mathrm{mas}$ (the Bright Star catalogue is limited to a distance of $\sim 200\, \mathrm{pc}$ for stars with $M_V=0$). After the removal of stars with spectral type later than B (according to the Bright Star catalogue classification) and of spectroscopic binaries, the sample was reduced to 516 stars. Note that the colours were deredenned using a procedure based on Str\"omgren $uvby\beta$ photometry \citep{napiwotzki}. The $uvby\beta$ photometry was obtained from the \citet*{hauck} catalogue. In Fig.~\ref{f:2} the colour-magnitude of sample~A is shown.

\subsection{Age and velocity determination}
The age of each individual star in sample~A was estimated using a method analogous to the one used to compute the evolutionary ages of the runaway stars (sample~B) (\citealp{silva1}, also see eg. \citealp{irrgang,ramspeck}). However, here the most recent evolutionary tracks by \citet{ekstrom} for non-rotating stars were used instead of the older models by the Geneva group. These tracks were computed using updated input physics and the revised value of the solar metallicity \citep*{asplund1}. The tracks were first converted from the $L-T_{\mathrm{eff}}$ to the $M_V-(B-V)_0$ space and then linearly interpolated in order to obtain an evolutionary age estimate. 

The conversion of the theoretical tracks was accomplished in two steps: the absolute magnitudes in $V$ band corresponding to the given luminosities were computed using the bolometric corrections by \citet*{flower1}; then the $(B-V)_0$ colours corresponding to the given effective temperatures were computed using the formulae in \citet{napiwotzki}. These converted tracks are shown in Fig.~\ref{f:2}.

The full space velocities were computed for all stars in both sample~A and sample~B were computed from the distances, measured proper motions and radial velocities. In the case of sample~A, the distances were derived from the respective {\it Hipparcos} parallaxes.

\subsection{Fitting procedure}
The method adopted to compute the pattern speed consisted in two steps: 1) tracing back the orbit of each star in order to determine its birthplace (assumed to lay on a spiral arm); 2) rotation of the position of the birthplace by the angle corresponding to the pattern speed $\Omega_\mathrm{p}$ and the age of the star. The best estimate of the pattern speed is the one that minimises the distance of the birthplaces to a given model of the spiral arms structure, for the sample. Note that other parameters of the model may be determined if a suitable sample is provided. This is essentially the method used by \citet{amaral1} and \citet{dias1} on their sample of open clusters. In our case we have adopted the cartographic model by \citet{vallee3} as a representation of the present spiral arms structure as it based on previous studies using a variety of tracers and it has a simple mathematical representation (logarithmic spiral). Thus, in polar coordinates:

\begin{equation}
r(\theta)=r_0 e^{b\theta+n\pi/2+\phi},
\label{eq:1}
\end{equation}

where $b=\tan p$, $\phi$ is the initial phase or phase angle, and $n=0,1,2,3$. The parameters of the spiral structure, as derived by \citet{vallee3}, are: the pitch angle $p=12.8^{\circ}$, the initial radius $r_0=2.1\ \mathrm{kpc}$, and the initial phase $\phi=-20^{\circ}$. However, these parameters were obtained assuming a distance of the Sun to the Galactic Centre, $R_{\odot}=7.6\ \mathrm{kpc}$, whereas in our study we assumed $R_{\odot}=8\ \mathrm{kpc}$. Taking this difference into account, a fit of the spiral structure to the tangent directions indicated in \citet{vallee3} gives an initial phase $\phi=-40^{\circ}\cdots-45^{\circ}$, for $R_{\odot}=8\ \mathrm{kpc}$ keeping the other parameters fixed. Given the nature of our samples (sample~A only includes local stars and sample~B is small), from these parameters only the initial phase $\phi$ was fitted together with the pattern speed $\Omega_\mathrm{p}$.

Thus, we used the following target function in our minimisation procedure:

\begin{equation}
{\cal F}(\Omega_\mathrm{p},\phi)=\sum_{\mathrm{i}}{\frac{1}{\sigma_\mathrm{x}(i)^2+\sigma_\mathrm{y}(i)^2}d_\mathrm{n}(i)},
\label{eq:2}
\end{equation}

where

\begin{equation}
d_\mathrm{n}(i)=\sqrt{\left(x_\mathrm{i}(\Omega_\mathrm{p},t_\mathrm{i})-x_\mathrm{s}(\theta_\mathrm{n},\phi)\right)^2+\left(y_\mathrm{i}(\Omega_\mathrm{p},t_\mathrm{i})-y_\mathrm{s}(\theta_\mathrm{n},\phi)\right)^2}
\label{eq:3}
\end{equation}

is the minimum distance between the birthplace rotated by $\Omega_\mathrm{p}$ (with coordinates $x_\mathrm{i},y_\mathrm{i}$) and the closest spiral arm (with the closest point having coordinates $x_\mathrm{s},y_\mathrm{s}$ corresponding to an angle $\theta_\mathrm{n}$), $t_\mathrm{i}$ is the estimated age of the star, and $\sigma_\mathrm{x}(i)$ and $\sigma_\mathrm{y}(i)$ are the uncertainties in the coordinates of the birthplaces as derived through a Monte Carlo error propagation procedure from the uncertainties in the input observable parameters. Since the fitting procedure assumes the spiral arms have no thickness (they are treated as lines) the uncertainty in the position of the stars was assumed to be a minimum of $200\ \mathrm{pc}$ (corresponding to a $3\, \sigma$ thickness of $1.2\ \mathrm{kpc}$, cf. \citealp{lepine1}; \citealp{russeil}), so the birthplaces far from the centre of the spiral arms are not excessively penalised by small weights during the minimisation procedure. This also alleviates the problem of stars born outside the spiral arms having a large influence in the fitting without resorting to some form of sigma-clipping as was done by \citet{naoz1} or \citet{dias1}, for example.

Since the target function, ${\cal F}(\Omega_\mathrm{p},\phi)$, does not strictly follow a $\chi^2$ distribution, and hence does not have known statistical properties, the error in the derived parameters was determined through a Monte Carlo procedure. Thus, 2000 replicas of the original sample were created, varying the input parameters (colour, parallax, proper motion and brightness) according to the respective errors distributions (assumed Gaussian). As new age estimates are computed in each realisation of the Monte Carlo procedure, this technique is also a good way of propagating the age uncertainties.

\section{Results and discussion}
\label{s:results}

\subsection{Pattern speed}
Although in principle it would be possible to constrain all the parameters of the simple logarithmic spiral model given by Equation~(\ref{eq:1}), in the concrete case of our samples this is not possible as the tracers used either cover only a very small part of the Galaxy (sample~A), or are too few with large measurement errors and small spread in ages (sample~B). Thus, only a fit to the phase angle, together with the pattern speed, was attempted.

\begin{figure}%
\includegraphics[width=82mm]{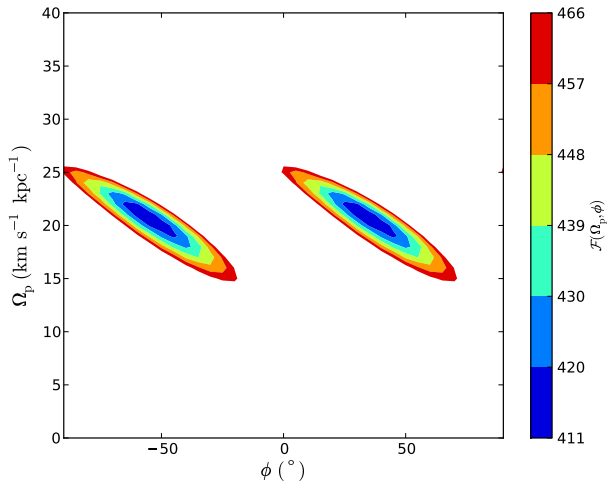}%
\caption{Contour plots of the $\chi^2$ values for sample~A stars.}%
\label{f:4}%
\end{figure}

\begin{figure}%
\includegraphics[width=82mm]{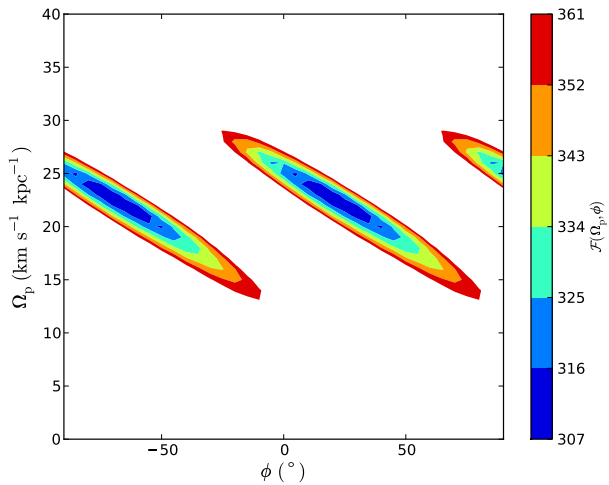}%
\caption{Contour plots of the $\chi^2$ values for sample~A stars older than $60\ \mathrm{Myr}$ (to partially remove the Gould Belt).}%
\label{f:44}%
\end{figure}

In order to look for the minimum of the target function (Equation~\ref{eq:2}) a grid spanning the area of interest of the parameter space was prepared in order to find the location of possible solutions which were later more precisely determined using an implementation of the Nelder-Mead simplex algorithm in {\sc python}\footnote{http://docs.scipy.org/doc/scipy/reference/generated/\\scipy.optimize.fmin.html}. In Fig.~\ref{f:4} the contour plot corresponding to sample-A is shown. Only one solution appears to exist (within the equivalent to $6\, \sigma$) which is repeated for phases separated by $90^{\circ}$, i.e. when one arm switches to the next one. However, care should be taken with contamination by stars formed in regions unrelated to the spiral arms, e.g. the Gould Belt. According to \citet{torra1}, the Gould Belt has an age of $\sim 60\ \mathrm{Myr}$, thus in order to remove this feature we computed a new grid considering only stars older than $\sim 60\ \mathrm{Myr}$. The resulting contour plot is shown in Fig.~\ref{f:44} and it is visible that although there is still only one solution it appears to become more uncertain.

\begin{figure}%
\includegraphics[width=82mm]{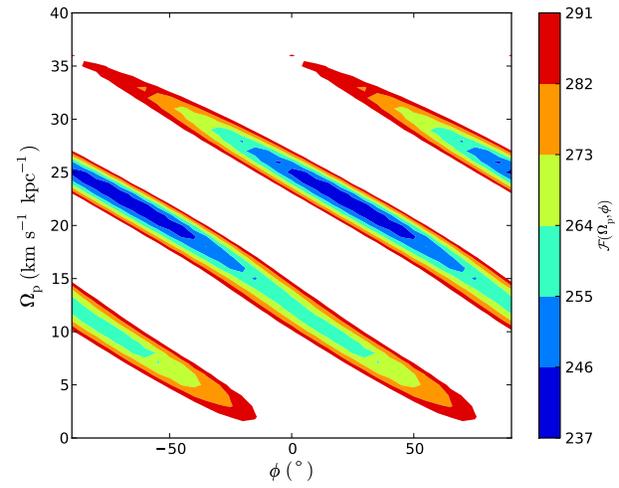}%
\caption{Contour plots of the $\chi^2$ values for sample~A stars older than $90\ \mathrm{Myr}$.}%
\label{f:5}%
\end{figure}

\begin{figure}%
\includegraphics[width=82mm]{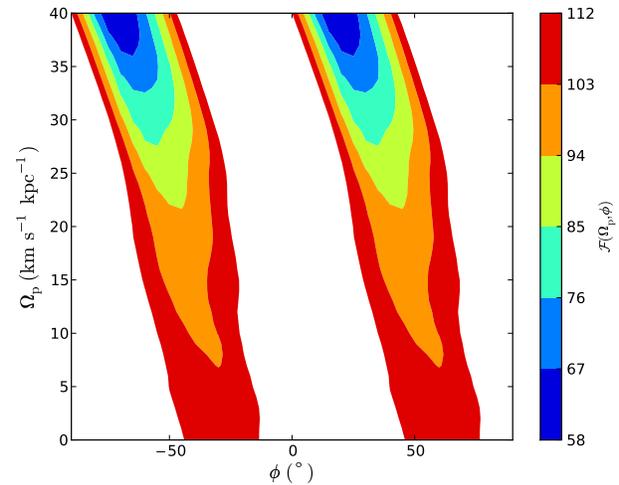}%
\caption{Contour plots of the $\chi^2$ values for sample~A stars younger than $60\ \mathrm{Myr}$.}%
\label{f:55}%
\end{figure}

The reason for this increased uncertainty is understood when the same procedure is applied to a sample of even older stars, as shown in Fig.~\ref{f:5}. In this case the solution becomes degenerate, showing that a sample of older stars does not provide a good constraint on the phase angle. On the other hand, the sample of stars younger than $60\ \mathrm{Myr}$, shown in Fig.~\ref{f:55}, provides a solution that behaves in the opposite way, providing a better constraint on the phase angle but performing worse in constraining the pattern speed. This is simply a reflection of the obvious fact that younger objects are good at describing the present state of the spiral arms but have not travelled enough to be able to carry information about the pattern speed, whereas the opposite is true for older objects.

\begin{figure}%
\includegraphics[width=82mm]{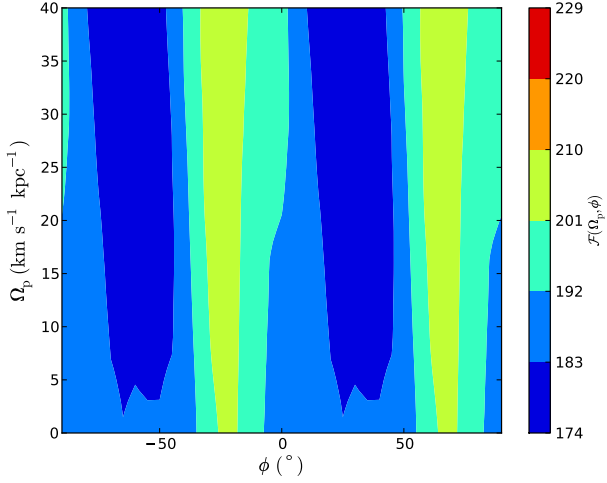}%
\caption{Phase angle estimate from sample of clusters younger than $7\ \mathrm{Myr}$ (from \citealp{dias2}).}%
\label{f:6}%
\end{figure}

\begin{figure}%
\includegraphics[width=82mm]{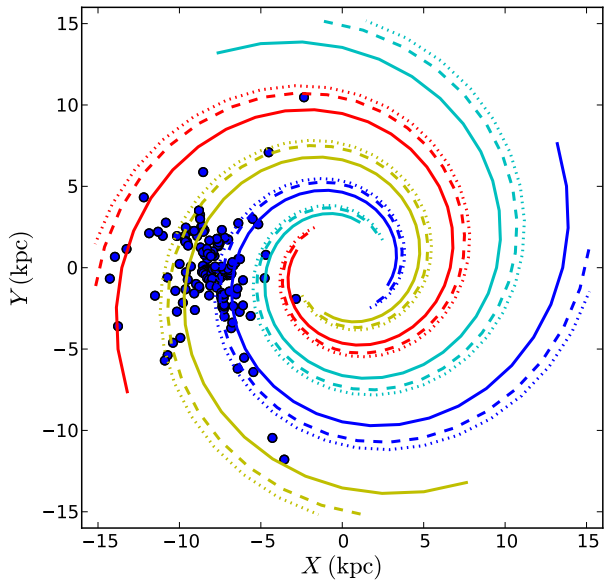}%
\caption{The spiral arms model by \citet{vallee3} with $\phi=-20^{\circ}$ (solid line) and $\phi=-45^{\circ}$ (dashed line), derived for $R_{\odot}=7.6\ \mathrm{kpc}$ and $R_{\odot}=8\ \mathrm{kpc}$ respectively, and $\phi=-55.5^{\circ}$ (dotted line). The different models are compared with the sample of open clusters younger than $<7\ \mathrm{Myr}$ (blue dots).}%
\label{f:7}%
\end{figure}

In fact, if the same procedure is applied to the young clusters ($<7\ \mathrm{Myr}$) of the \citet{dias2} sample, we find that the present phase angle appears closer to $-50^{\circ}$ than to the $-20^{\circ}$ derived by \citet{vallee3}. However, as we have seen, most of this difference is attributable to a difference in the chosen distance of the Sun to the Galactic Centre ($R_{\odot}=8\ \mathrm{kpc}$ in our case), and, in particular, the constrains used by \citet{vallee3} imply $\phi\simeq -45^{\circ}$ assuming $R_{\odot}=8\ \mathrm{kpc}$. In Fig.~\ref{f:6} we have plotted the grid of the parameter space for this sample of clusters, and in Fig.~\ref{f:7} a comparison between the two different angles and how they compare with the sample of young clusters. Although a phase angle of $-20^{\circ}$ appears to be a better fit in some regions, a larger angle fits better other regions (Perseus arm, for example). Even though the agreement between our solution and the one obtained by \citet{vallee3} (and hence with constraints provided by the tangential directions and the distance to the Perseus arm obtained by \citealp{xu1}) is not perfect it would be possible to find a solution consistent with the constraint, however that would require further analysis beyond the scope of this paper. Note that in the plot shown in Fig.~\ref{f:6} a phase angle of $-20^{\circ}$ is actually in the region of the worst solutions, but a phase angle of $-45^{\circ}$ (the corrected value for $R_{\odot}=8\ \mathrm{kpc}$) is in the region of best solutions.

\begin{table}
\caption{Pattern speed ($\Omega_\mathrm{p}$) and phase angle ($\phi$) obtained for three different samples: entire sample~A (local {\it Hipparcos} stars), only stars older than $60\ \mathrm{Myr}$ in sample~A, sample~B (runaway stars).}
\label{tab:1}
\centering
\begin{tabular}{@{}lcc}
\hline
Sample & $\Omega_\mathrm{p}\ (\mathrm{km}\, \mathrm{s}^{-1}\, \mathrm{kpc}^{-1})$ & $\phi\ (^{\circ})$\\
\hline
A (all) & $20.3\pm 0.5$ & $-53.1\pm 3.1$\medskip\\

A ($>60\ \mathrm{Myr}$) & $21.2\pm 1.1$ & $-60.3\pm 7.4$\medskip\\

B & $21.9\pm 8.6$ & $-63.5\pm 22.5$\\
\hline 
\end{tabular}
\end{table}

The minimisation of the target function (Equation~\ref{eq:2}) around the solution found in the parameter space, $(\Omega_\mathrm{p}=20,\phi=-50^{\circ})$, yielded the values seen in Table~\ref{tab:1} for three different samples: all of sample~A, only stars in sample~A that are older than $60\ \mathrm{Myr}$, and the sample~B. Note that the solution in the case of sample~B is quite degenerate, however there is a solution compatible with the one found in the other two cases.

\begin{figure}%
\includegraphics[width=82mm]{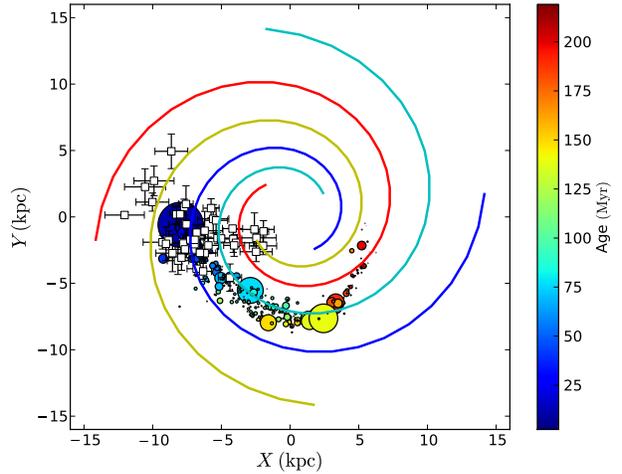}%
\caption{Distribution of the birthplaces in their present positions, resulting from the derived pattern speed. Sample~A and sample~B (runaway stars) are plotted as coloured circles (with sizes proportional to the uncertainty in the position) and squares with error bars, respectively. Only runaway stars with standard deviations $< 1.5\ \mathrm{kpc}$ are shown. The spiral arms are also plotted in their current position}%
\label{f:8}%
\end{figure}

The spiral arm structure with the phase estimated from sample~A is shown in Fig.~\ref{f:8}, together with the birthplaces of stars from samples A and B rotated from their original positions by the derived (from sample~A) pattern speed times the ages of the stars. It is interesting to note how the runaways seem to have been born in many different arms, and in particular how well they appear to follow the ``yellow'' arm (to be identified with the Perseus arm).

\subsection{Comparison with other studies}
Estimates of the pattern speed in the literature cover a wide range of values: using different variations of the ``birthplace'' determination method with samples of open clusters, \citet{amaral1} found $\Omega_\mathrm{p}\approx 20\ \mathrm{km}\, \mathrm{s}^{-1}\, \mathrm{kpc}^{-1}$ and \citet{dias1} $\Omega_\mathrm{p}\approx 24\pm 1\ \mathrm{km}\, \mathrm{s}^{-1}\, \mathrm{kpc}^{-1}$; by fitting a kinematical model of the Milky Way \citet{fernandez1} found $\Omega_\mathrm{p}\approx 30\ \mathrm{km}\, \mathrm{s}^{-1}\, \mathrm{kpc}^{-1}$ (with an uncertainty between $2\ \mathrm{km}\, \mathrm{s}^{-1}\, \mathrm{kpc}^{-1}$ and about $7\ \mathrm{km}\, \mathrm{s}^{-1}\, \mathrm{kpc}^{-1}$ depending on the specific tracer); \citet{martos1} obtained $\Omega_\mathrm{p}\approx 20\ \mathrm{km}\, \mathrm{s}^{-1}\, \mathrm{kpc}^{-1}$ by fitting dust observations with a dynamical model of the Galaxy. Moreover, \citet{naoz1} estimate different pattern speeds, using a variation of the ``birthplace'' method with a sample of open clusters, for different structures (with an uncertainty close to $\sim 2\ \mathrm{km}\, \mathrm{s}^{-1}\, \mathrm{kpc}^{-1}$ for all estimates): the Sagittarius-Carina arm is actually a superposition of two spiral structures with $\Omega_\mathrm{p,1}\approx 16.5\ \mathrm{km}\, \mathrm{s}^{-1}\, \mathrm{kpc}^{-1}$ and $\Omega_\mathrm{p,2}\approx 29.8\ \mathrm{km}\, \mathrm{s}^{-1}\, \mathrm{kpc}^{-1}$, the Perseus arm with $\Omega_\mathrm{p}\approx 20\ \mathrm{km}\, \mathrm{s}^{-1}\, \mathrm{kpc}^{-1}$, and the Orion ``armlet'' with $\Omega_\mathrm{p}\approx 28.9\ \mathrm{km}\, \mathrm{s}^{-1}\, \mathrm{kpc}^{-1}$. By fitting models of gas flow in the Milky Way to CO observations, \citet{bissantz1} obtained $\Omega_\mathrm{p}\approx 20\ \mathrm{km}\, \mathrm{s}^{-1}\, \mathrm{kpc}^{-1}$.

The different estimates of the pattern speed seem to suggest (even though it is not clear how uncertain some of these estimates are) a value in the range $\Omega_\mathrm{p} \sim 20 - 25\ \mathrm{km}\, \mathrm{s}^{-1}\, \mathrm{kpc}^{-1}$ (see also \citealp{gerhard1}), which would be compatible with our own estimates. Furthermore, there is independent confirmation, from the observation of the predicted ring-shaped gap in the radial $\mathrm{H}\,\mathrm{I}$ density distribution \citep{amores1}, of a corotation radius close to the value of the distance of the Sun to the Galactic centre, implying a pattern speed $\Omega_\mathrm{p} \sim 25 - 29\ \mathrm{km}\, \mathrm{s}^{-1}\, \mathrm{kpc}^{-1}$. As we have seen, our best estimate gives a value of $\Omega_\mathrm{p} = 20.3\pm 0.5\ \mathrm{km}\, \mathrm{s}^{-1}\, \mathrm{kpc}^{-1}$ for the pattern speed, which, although consistent with other determinations, appears to be at odds with estimates of the corotation radius, particularly given the small uncertainty. However, we must bear in mind certain caveats to this result: the uncertainty includes only the formal statistical contributions, ignoring systematic effects; it is possible that the inclusion of regions of star formation unrelated with the spiral arms may introduce biases in the estimates. Indeed, it can be seen in Table~\ref{tab:1} that removing most of the Gould Belt from the sample produces an estimate ($\Omega_\mathrm{p} = 21.2\pm 1.1\ \mathrm{km}\, \mathrm{s}^{-1}\, \mathrm{kpc}^{-1}$) closer to the upper limit of the suggested range for the pattern speed ($\Omega_\mathrm{p} \sim 20 - 25\ \mathrm{km}\, \mathrm{s}^{-1}\, \mathrm{kpc}^{-1}$), albeit with a larger uncertainty caused by the loss of the young stars constraining the current position of the arms.

For the same reason, it is possible that the uncertainty on the phase angle determination was underestimated, because the large number of young stars in the solar neighbourhood may introduce a bias in the position of the spiral arms, as they might have been formed in other regions (like the Gould Belt, or interarm features). However, the values we obtained for the phase angle are close to the estimates by \citet{fernandez1} (varying between $\phi=-45^{\circ}\pm 32^{\circ}$ and $\phi=8^{\circ}\pm 52^{\circ}$, when assuming models with four spiral arms), and also with the estimate obtained from the sample of open clusters (see Fig.~\ref{f:7}).

\section{Prospects with {\it Gaia}}
The {\it Gaia} mission will deliver trigonometric parallaxes, improved proper motions and spectral energy distributions for many early-type runaway stars. This will: 1) improve the achievable accuracy for the existing sample, and 2) allow the selection of a much increased sample of runaway stars.

In order to illustrate the expected increase in performance of our method once {\it Gaia} data becomes available, we have determined the birthplaces of two representative stars of our sample of runaways, one far away and one nearby, assuming the knowledge of their parallaxes and proper motions with the predicted {\it Gaia} accuracy. The results obtained are summarised in Table~\ref{tab:3} and contrasted with the previous determination. Our results for the nearby star HIP~59955 are already fairly accurate. {\it Gaia} measurements will reduce the overall uncertainty of the determination of the birthplace to a level comparable with the typical width of a spiral arm. On the other hand, the improvement for the star EC~04420-1908 is dramatic, with the $X$ coordinate uncertainty decreasing by $\sim 500$~per~cent and the $Y$ coordinate uncertainty decreasing by $\sim 300$~per~cent. Although the uncertainty obtained in this case is larger than the typical width of a spiral arm, it is now smaller than the expected interarm distance \citep{vallee3}. 

\begin{figure}%
\includegraphics[width=82mm]{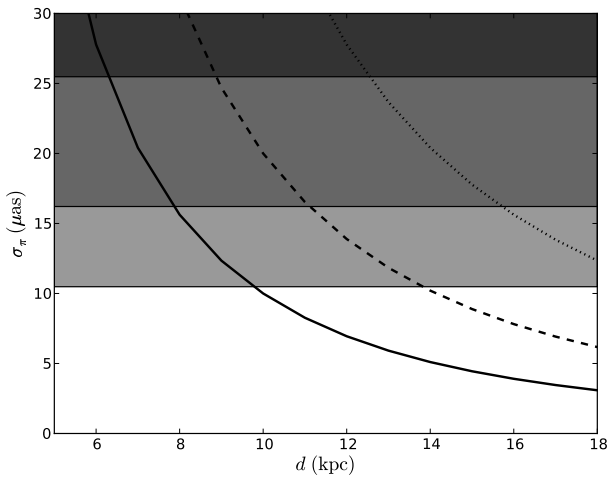}%
\caption{Parallax error as a function of distance corresponding to an error of $1\ \mathrm{kpc}$ (solid line), $2\ \mathrm{kpc}$ (dashed line) and $3\ \mathrm{kpc}$ (dotted line) in distance. The shaded areas correspond to the errors for stars with $G<15$ (dark grey), $G<14$ (medium grey) and $G<13$ (light grey), that are predicted for {\it Gaia}.}%
\label{f:13}%
\end{figure}

Noting that the distance uncertainty that would be obtained from the {\it Gaia} parallax in the case of EC~04420-1908 is $\sim 1\ \mathrm{kpc}$, it is clear that this error should be considered an upper limit if we hope to obtain precise estimates of the birthplaces' coordinates. The parallax errors that correspond to an error of $1\ \mathrm{kpc}$ in the distance, as a function of distance, are plotted in Fig.~\ref{f:13}, together with the parallax errors for stars with $V<13$ and $V<14$ (according to the formula in the {\it Gaia} Science Performance website\footnote{http://www.rssd.esa.int/index.php?project=GAIA\&\\page=Science\_Performance}). It is possible to conclude from this plot that it will be possible to determine directly -- with high precision -- the birthplaces of stars with $V<13$ and $V<14$ as far as $10\ \mathrm{kpc}$ and $8\ \mathrm{kpc}$, respectively. For stars further away other methods will be needed in order to attain a high precision, however it is important to note that distances of $14\ \mathrm{kpc}$ and $12\ \mathrm{kpc}$, respectively for stars brighter than $V=13$ and $V=14$, will provide parallax errors of $\approx 15$~per~cent.

\begin{table}
\caption{Precision of the determination of the birthplaces of the stars EC~04420-1908 and HIP~59955 using present data \citep{silva1} and the precision predicted using {\it Gaia} data. Estimates of the distances and the coordinates of the birthplaces obtained using present data are indicated with a p subscript whereas the estimates obtained using {\it Gaia} precision data are indicated with a g subscript.}
\label{tab:3}
\centering
\begin{tabular}{@{}lrr}
\hline
 & EC~04420-1908 & HIP~59955\\
\hline
$V$ & $13$ & $9.71$\\
 & & \\
$d_\mathrm{p}\ \mathrm{(kpc)}$ & $10.44^{+4.35}_{-3.16}$ & $1.5^{+0.38}_{-0.30}$\\
$d_\mathrm{g}\ \mathrm{(kpc)}$ & $10.55\pm 1.2$ & $1.5\pm 0.02$\\
 & & \\
$X_\mathrm{p}\ \mathrm{(kpc)}$ & $-11.43^{+4.13}_{-4.97}$ & $-4.31^{+1.37}_{-1.11}$\\
$X_\mathrm{g}\ \mathrm{(kpc)}$ & $-12.14\pm 0.88$ & $-4.30\pm 0.60$\\
 & & \\
$Y_\mathrm{p}\ \mathrm{(kpc)}$ & $-7.40^{+2.74}_{-4.29}$ & $-4.29^{+0.38}_{-0.37}$\\
$Y_\mathrm{g}\ \mathrm{(kpc)}$ & $-8.22\pm 0.78$ & $-4.39\pm 0.34$\\\hline
\end{tabular}
\end{table}

\subsection{Sample selection and contamination with old evolved stars}
An issue always present in studies of runaway stars is the contamination by low mass stars in advanced stages of evolution, in particular the post-AGB and core Helium burning Horizontal Branch, as they may mimic the atmospheric parameters of main sequence stars (\citealp{tobin1}; \citealp{martin1}). However, as a direct measurement of the distance will be available with {\it Gaia} it will be possible to select runaway stars based on their absolute magnitudes $M_V$ and effective temperatures $T_{\mathrm{eff}}$.

From Fig.~\ref{f:13}, we have already seen that it will possible to determine the distance with high precision for stars with $G<13$, and, more generally, with an error $<15$~per~cent for stars with $G<14$, at least for distances up to $12\ \mathrm{kpc}$. However, it can also be seen that an error $<30$~per~cent will be possible for stars with $G<15$. Thus, considering that this error in the distance translates to an error of 0.65 in the absolute brightness $M_V$, it will be possible to use {\it Gaia} parallaxes to select a sample of early type runaway stars, even in those cases where the precision is not good enough to be used directly, as can be seen in Fig.~\ref{f:9}. This is because the absolute magnitudes of stars with spectral type B7 or earlier (effective temperature $\sim 13500\ \mathrm{K}$) are much lower than the absolute magnitude of the Terminal Age Horizontal Branch (TAHB, the end of Helium burning in the core), which is $M_V>1$, for the same temperature. Note that this allows the selection of B7 stars as far as $12\ \mathrm{kpc}$.

\begin{figure}%
\includegraphics[width=82mm]{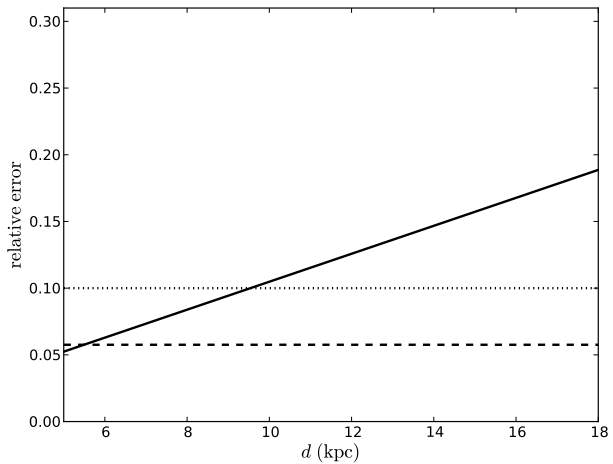}%
\caption{Comparison between the expected relative error, as a function of distance, for the distance estimates: from {\it Gaia} parallaxes for a $G<13$ star (solid line), from spectroscopy when the uncertainty in $\log g$ is $0.05\ \mathrm{dex}$ (dashed line), as obtained by \citet{nieva1}, taking all error sources into account (dotted line).}%
\label{f:14}%
\end{figure}

\begin{figure}%
\includegraphics[width=82mm]{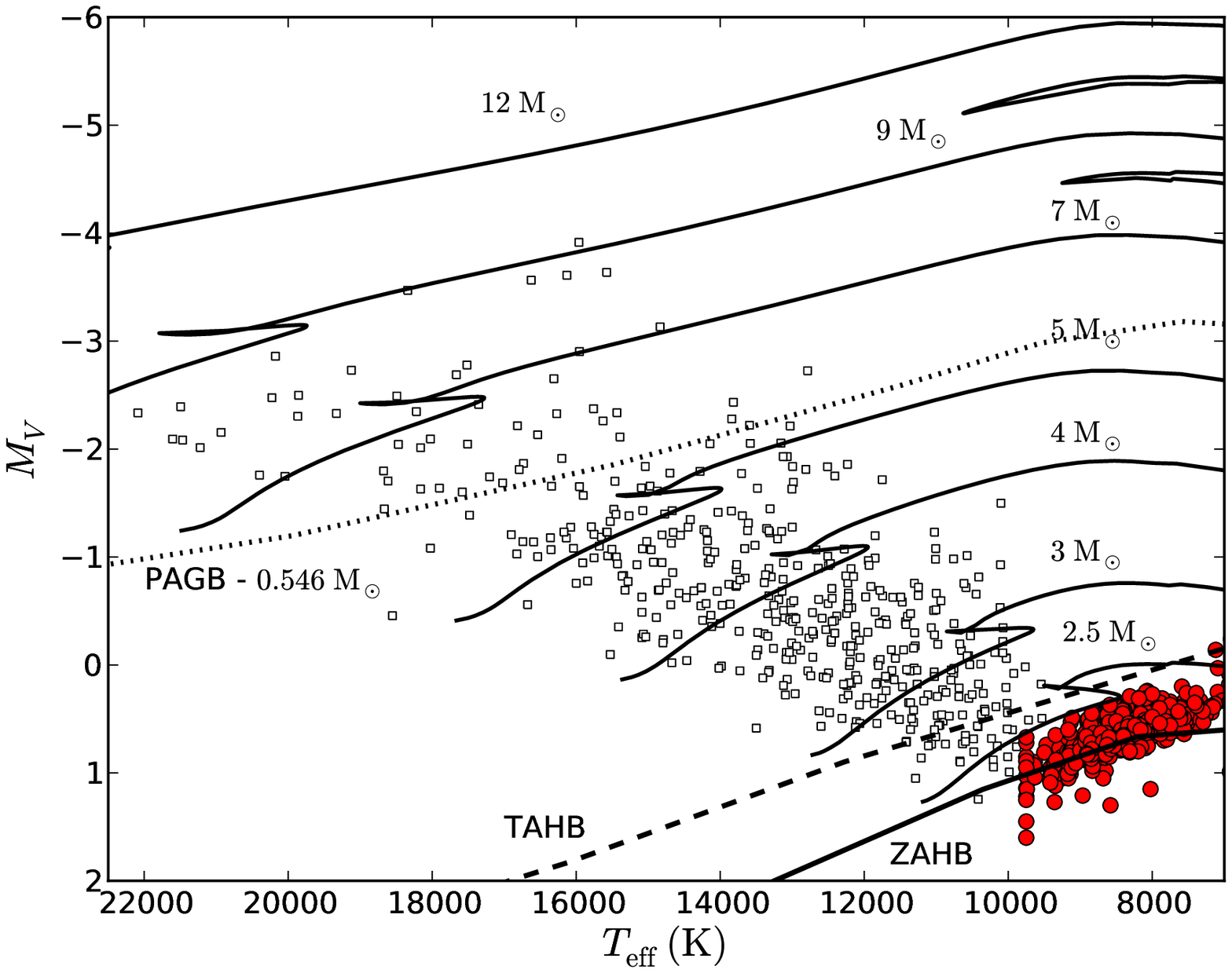}%
\caption{The Zero Age Horizontal Branch (ZAHB) and the Terminal Age Horizontal Branch (TAHB), from the models by \citet{dorman} for stars with an Helium mass fraction $Y=0.247$ and a metallicity $[\mathrm{Fe}/\mathrm{H}]=-1.48$, and the theoretical evolutionary track of a low mass Post-AGB star, from the models by \citet{schoenberner}, in a $M_V-T_{\mathrm{eff}}$ diagram. The absolute magnitudes for the Horizontal Branch models were obtained by applying the bolometric corrections of \citet{flower1}. The red dots correspond to stars identified as Horizontal Branch in the survey by \citet{brown1}.}%
\label{f:9}%
\end{figure}

Once the sample is selected, better estimates of the distance could be obtained after a careful spectroscopic analysis in the case of the distant stars. Since the limiting factor of spectroscopic distance estimates is usually the uncertainty in the $\log g$ estimate, it is instructive to see how the error that is expected from {\it Gaia} parallaxes compares with good spectroscopic estimates, with small uncertainties in $\log g$. For example, \citet{nieva1} obtain a distance accuracy of $\sim 10$~per~cent in their study of nearby B stars, from an analysis where the uncertainty in the $\log g$ determination is $\sim 0.05\ \mathrm{dex}$. The comparison between these estimates, and also the expected error if only the $\log g$ uncertainty (of $\sim 0.05\ \mathrm{dex}$) is considered, is shown in Fig.~\ref{f:14}. Note how the {\it Gaia} estimates have acceptable errors only up to distances of $10$ to $12\ \mathrm{kpc}$. Moreover, since the radial velocities measured with {\it Gaia} will have acceptable errors ($< 10\ \mathrm{km}\, \mathrm{s}^{-1}$) only in the case of the brightest stars ($V=13\ldots 14$), follow-up spectroscopy may be needed in order to obtain good estimates of the radial velocities of the more distant stars.

Also important, for both the selection of runaway stars and the follow-up spectroscopic analysis in the relevant cases, are good estimates of the effective temperatures. These could be obtained from {\it Gaia} spectrophotometry, if we assume it will be at least as accurate for effective temperature determination as medium band photometry, e.g. Str\"omgren-Crawford $uvby\beta$ photometry. The expected error would be then, according to \citet{napiwotzki}, about 3~per~cent in the range $11000\ \mathrm{K}\leq T_{\mathrm{eff}} \leq 20000\ \mathrm{K}$ and 4~per~cent for hotter stars. These should be conservative estimates of the errors, which could be further refined by adding {\it Galex} UV photometric data \citep*{martin_galex}.

As was already suggested, the expected contamination by Horizontal Branch stars, for stars with spectral type B7 or earlier will be extremely low, if the appropriate selection criteria is adopted. Adopting a criteria similar to the one used by \citet{silva1} (dependent on estimates of $\log g$ obtained from spectroscopy) should limit the contamination to less than 10~per~cent of the total. In fact, most contamination by Horizontal Branch stars can be eliminated by applying a cut in the $T_{\mathrm{eff}}-M_V$ space (e.g. $T_{\mathrm{eff}}>13500$ and $M_V<0$), as can be seen in Fig.~\ref{f:9}. However, this can be improved by combining good distance estimates with $\log g$ information, since the distances of $0.5\ \mathrm{M}_\odot$ and $2.5\ \mathrm{M}_\odot$ stars of the same observed brightness differ by $\sqrt{5}$. The contamination by post-AGB stars is more difficult to ascertain, however these stars are very rare and usually account for a very small amount of contamination in studies of runaway stars (e.g. $\sim 8$~per~cent in the study by \citealp{martin2}, and $\sim 5$~per~cent in the study by \citealp{silva1}). Nevertheless, the proposed criteria can be applied to these stars and also to stars in other post-Horizontal Branch evolutionary stages.

In summary, combined with a small proportion of Horizontal Branch stars evolved past the TAHB, the level of contamination with old evolved stars should be $\lesssim 10$~per~cent. Furthermore, we will be able to use {\it Gaia} astrometry to select runaway stars with spectral types earlier than B7 up to a distance $\sim 14\ \mathrm{kpc}$ and spectral types earlier than B3 up to a distance $\sim 22\ \mathrm{kpc}$, assuming a brightness $V=15$. However, follow-up spectroscopic analysis will be needed to obtain good estimates of the distances in the case of the fainter stars. On the other hand, for stars brighter than $V=13$ (corresponding to a distance $\sim 9\ \mathrm{kpc}$ for B3 and $\sim 5\ \mathrm{kpc}$ for B7 stars) it will be possible to obtain good distance estimates directly from {\it Gaia} parallaxes.

\subsection{Spiral arms on the other side of the Galaxy}
As we have seen, it will be possible to select a sample of runaway stars reaching distances up to $\sim 14\ \mathrm{kpc}$, creating the opportunity to study the spiral arms behind the centre of our Galaxy. It is thus important to have an estimate of how many runaway stars will be observed behind the Galactic centre. The estimated number density of runaway stars high above the Galactic plane is $\sim 9\ \mathrm{kpc}^{-2}$ \citep{silva1}. This corresponds to a total of $\sim 6400$ stars in the Galaxy, in a $15\ \mathrm{kpc}$ radius (see also \citealp{napiwotzki2}). Of these, about one third are at a distance $>3\ \mathrm{kpc}$ from the Galactic plane. 

From the reddening maps of \citet{schlegel1} it is possible to estimate the expected extinction for stars outside the Galactic plane, and in particular to obtain an upper limit to the expected extinction for runaway stars high above the Galactic plane, considering that the greatest amount of extinction occurs in the direction of the Galactic centre ($l=0^{\circ}$). The extinction for stars $3\ \mathrm{kpc}$ away from the Galactic plane depends on the distance under consideration, as the Galactic latitude $b$ decreases for larger distances. In Table~\ref{tab:2} we show the values of extinction $A_V$ obtained from the Schlegel maps for distances in the range $10-16\ \mathrm{kpc}$. A distance of $10\ \mathrm{kpc}$ should be enough to probe the Sagittarius-Carina arm, $12\ \mathrm{kpc}$ should probe the Perseus arm, $14\ \mathrm{kpc}$ should probe the Cygnus arm, $16\ \mathrm{kpc}$ should probe the Crux-Scutum arm, and should be close to the corotation radius on the other side of the Galaxy. We have included the expected extinction in the Northern and Southern Galactic hemispheres separately because they differ by a large margin due to the presence of the Rho Ophiuchi cloud complex in the Northern hemisphere. Also included in Table~\ref{tab:2} are the apparent magnitudes of B3 and B7 (reddened) stars at the indicated distances. It should be noted, however, that these reddening estimates are actually overestimates, since {\it Gaia} $G$ is redder than Johnson-$V$ making it less sensitive to interstellar absorption.

For a B star with $V<15$, we can infer from Table~\ref{tab:2} that we will be able to observe stars with spectral type earlier than B3 as far as $12\ \mathrm{kpc}$ in both hemispheres and as far as $14\ \mathrm{kpc}$ in the Southern hemisphere. On the other hand, stars with later spectral types (but earlier than B7) will be observable only in the Southern hemisphere but as far as $12\ \mathrm{kpc}$. However, it will be possible to use B3 stars (and earlier) to probe distance as far as $18\ \mathrm{kpc}$ in the Southern hemisphere and $14\ \mathrm{kpc}$ in the Northern hemisphere, although with a loss of accuracy. In the same manner, it will be possible to use B7 stars (and earlier) as far as $14\ \mathrm{kpc}$ in the Southern hemisphere and as far as $12\ \mathrm{kpc}$ in the Northern hemisphere, although with less accuracy.

However, we should remember that we are limited to distances of $\sim 5\ \mathrm{kpc}$ and $\sim 9\ \mathrm{kpc}$ for B3 and B7 stars, respectively, in terms of accurate distances determined from {\it Gaia} parallaxes, as was previously explained. Nevertheless, as was also pointed out, it should be possible to obtain accurate distances from follow-up spectroscopic analysis but this is within easy reach of modern medium-large telescopes. It is also important to note that this does not invalidate that, at least for a fraction of the sample (i.e. stars with spectral types earlier than B3), it will be possible to use direct distance determinations from {\it Gaia}.

\begin{table}
\caption{Extinction predicted by the \citet{schlegel1} reddening maps for different Galactic latitudes. The Galactic latitudes were computed for points $3\ \mathrm{kpc}$ above the Galactic plane and at the given distances ($d$). Also indicated are the apparent magnitudes for stars with spectral types B3 and B7 when reddened by the corresponding amounts. The values for the Northern and Southern Galactic hemispheres are distinguished by the use of parentheses, with the values inside parentheses corresponding to the Southern hemisphere.}
\label{tab:2}
\centering
\begin{tabular}{@{}lcccc}
\hline
$d\ \mathrm{(kpc)}$ & $|b|\ (^{\circ})$ & $A_V$ & $V\ \mathrm{(B3)}$ & $V\ \mathrm{(B7)}$\\
\hline
10 & $17$ & $1.7\ (0.4)$ & $15.6\ (14.3)$ & $16.7\ (15.4)$\\
12 & $14$ & $1.0\ (0.4)$ & $15.3\ (14.7)$ & $16.4\ (15.8)$\\
14 & $12$ & $1.4\ (0.5)$ & $16.0\ (15.1)$ & $17.1\ (16.2)$\\
16 & $11$ & $2.0\ (0.4)$ & $16.9\ (15.3)$ & $18.0\ (16.4)$\\
\hline 
\end{tabular}
\end{table}

Thus, from the expected sample of $\sim 2100$ runaway stars $3\ \mathrm{kpc}$ above the Galactic plane, 525 are expected to be within $45^{\circ}$ of the direction of the Galactic centre, if they are uniformly distributed. Hence, given the previous considerations, we expect to have a sample of $\sim 250$ in the Southern hemisphere covering a distance up to $18\ \mathrm{kpc}$ and $\sim 175$ in the Northern hemisphere  covering a distance up to $14\ \mathrm{kpc}$. The total number of stars should then be close to 400.

\section{Conclusions}
\label{s:conclusions}
The determination of the birthplaces of early-type stars is a feasible method to trace the position of the spiral arms in different time instants. We were able to estimate the pattern speed and the (present) phase angle of the Galaxy's spiral structure from a sample of local stars and obtained values ($\Omega_\mathrm{p} = 20.3\pm 0.5\ \mathrm{km}\, \mathrm{s}^{-1}\, \mathrm{kpc}^{-1}$ and $\phi=-53.1^{\circ}\pm 3.1^{\circ}$) that are consistent with previous estimates. Moreover, although these estimates are sensitive to systematic effects, in particular biases introduced by stars born outside the spiral arms, these effects are not crippling and introduce an extra error of $\sim 10$~per~cent.

From the sample of high Galactic latitude runaway stars (sample~B) we also derived a consistent estimate of the parameters ($\Omega_\mathrm{p} = 21.9\pm 8.6\ \mathrm{km}\, \mathrm{s}^{-1}\, \mathrm{kpc}^{-1}$ and $\phi=-63.5^{\circ}\pm 22.5^{\circ}$). However, given the nature of this sample (larger observational errors, small numbers and small spread in ages) the solution is not as well defined as in the case of the one obtained from sample~A and has thus a larger uncertainty.

More importantly, we have shown that this method has the potential to be used with data obtained by {\it Gaia} to retrieve the birthplaces of distant runaway stars and hence it will be possible to trace not only the closest spiral arms but also the more distant ones, in particular the portions behind the Galactic centre, which are usually obscured by interstellar reddening. We have estimated that the number of runaway stars that will be available to trace these distant portions of the spiral arms, in particular the far sides of the Sagittarius-Carina and Perseus arms, will be $\sim 250$ to $\sim 400$, whereas the number of runaway stars available to trace spiral arms in closer regions of the Galaxy will be $> 1000$. Thus, considering the availability of the high quality astrometric data obtained by {\it Gaia} in the near future, it will be possible to apply the method to a large number of runaway stars with high accuracy.

\section*{Acknowledgments}
M.S. gratefully acknowledges financial support by the University of Hertfordshire. The authors wish to thank the referee for helpful comments that helped to improve the paper.
This research has made use of NASA's Astrophysics Data System and of the SIMBAD database, operated at CDS, Strasbourg, France.

\bibliographystyle{mn2e}
\bibliography{refs_spiral}

\end{document}